\documentclass[journal,twoside,web]{bin/ieeecolor}
\usepackage{bin/generic}
\usepackage{cite}
\usepackage{amsmath,amssymb,amsfonts}
\usepackage{algorithmic}
\usepackage{graphicx}
\usepackage{textcomp}
\usepackage{soul}

\makeatletter

\def\fnum@figure{\textcolor{subsectioncolor}{\sf Fig.~\thefigure}}
\def\fnum@table{\textcolor{subsectioncolor}{\sf TABLE~\thetable}}

\def\ps@headings{\def\@oddhead{}\def\@evenhead{}\def\@oddfoot{\hfil\thepage\hfil}\def\@evenfoot{\hfil\thepage\hfil}}
\def\ps@titlepagestyle{\def\@oddhead{}\def\@evenhead{}\def\@oddfoot{\hfil\thepage\hfil}\def\@evenfoot{\hfil\thepage\hfil}}
\makeatother

\begin{document}
\title{SH-SAW Acousto-Electric Amplifier in Epitaxial InGaAs on Lithium Niobate on Insulator}
\author{Chuan Tian, Christopher Heidelberger, and Siddhartha Ghosh
\thanks{Chuan Tian and Siddhartha Ghosh are with Northeastern University, Boston, MA, USA.}
\thanks{Christopher Heidelberger is with MIT Lincoln Laboratory, Lexington, MA, USA.}}
\maketitle
\begin{abstract}
This work demonstrates shear-horizontal surface acoustic wave (SH-SAW) acoustoelectric (AE) amplification on an epitaxial InGaAs / X-cut lithium niobate on insulator (LNOI) heterostructure formed by Al$_2$O$_3$-mediated wafer bonding. Deployable passivated devices show a stable fundamental-mode non-reciprocity of 32~dB/mm at 1.11~GHz (30~V bias, 64~mW consumed), while unpassivated devices reach 174~dB/mm across 1.1--2.8~GHz, reported as upper bounds. Device characterization establishes the role of mode-dependent \textit{K}$^2$ in determining the achievable gain. Hall-effect measurements of the transferred InGaAs serve as a quantitative diagnostic: the extracted carrier density, elevated by unintentional silicon doping during epitaxy, accounts for the absolute AE gain when inserted into the analytical model and identifies epitaxial process control as a clear lever for further enhancement. We further identify ambient oxidation of the bare InGaAs surface as a distinct aging mechanism that extinguishes the AE response within weeks, and show that an InP or ALD Al$_2$O$_3$ passivation layer suppresses it, at the cost of redistributing the piezoelectric field away from the channel. These results establish InGaAs-on-LNOI as a compact, low-power platform for non-reciprocal RF components and acoustoelectric delay lines, with strong relevance to in-band full-duplex (IBFD) transceivers and spectrum-efficient wireless front ends.
\end{abstract}

\begin{IEEEkeywords}
acoustoelectric effect, surface acoustic wave (SAW), lithium niobate on insulator (LNOI), InGaAs, heterogeneous integration, non-reciprocal RF devices
\end{IEEEkeywords}
\section{Introduction}

\begin{figure*}
\centering
\includegraphics[width = 1\textwidth]{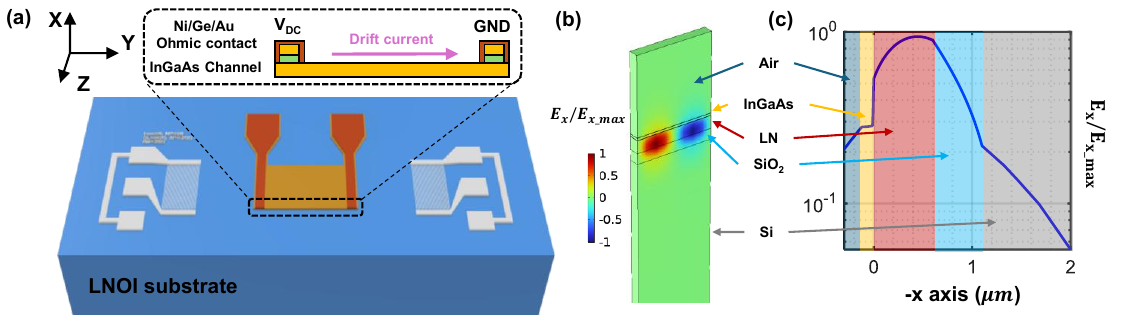}
\caption{\label{fig:schem}Device architecture and simulated electric-field distribution of the acoustoelectric amplifier. (a) Schematic cross-section of the SH-SAW device integrating a bonded InGaAs semiconductor channel on an LNOI substrate with Ni/Ge/Au ohmic contacts. (b) Simulated lateral electric-field distribution of the SH-SAW mode at a wavelength of $3~\mu\mathrm{m}$, normalized to the maximum value ($E_x/E_{x,\max}$). The InGaAs film is modeled with a finite electrical conductivity of $420~\mathrm{S/m}$, corresponding to an ideal intrinsic InGaAs thin film (carrier concentration 3$\times10^{15}cm^{-3}$ and mobility 8740 $cm^2/V\cdot s$), which introduces partial electrostatic screening of the piezoelectrically induced electric field. (c) Normalized electric-field profile ($E_x/E_{x,\max}$) extracted along the vertical cutline, illustrating the field penetration across the air, InGaAs, lithium niobate, buried oxide layers and silicon substrate
}
\end{figure*}

Modern radio-frequency (RF) systems increasingly require compact, energy-efficient, and broadband signal-processing components to support rapidly expanding data volumes. A key feature envisioned for fifth-generation (5G) and future wireless networks is in-band full duplex (IBFD) communication, which enables simultaneous transmission and reception within the same frequency band and thereby improves spectrum utilization, reduces architectural complexity, and enhances overall system efficiency \cite{kenn_ibfd_book}. Beyond commercial networks, IBFD is also relevant to defense applications such as jammer systems, where simultaneously suppressing adversary communication and detecting weak incoming signals is essential.

A central challenge in IBFD systems is self-interference cancellation (SIC), which must sufficiently suppress the strong transmit signal at the receiver. One promising time-domain SIC approach splits the transmit waveform into multiple delayed taps that are individually weighted and recombined to destructively interfere with self-interference. Such architectures require wide-range, high-fidelity true-time-delay (TTD) synthesis with large time–bandwidth products (TBP), as well as the ability to implement weighted or reconfigurable taps. Realizing these functionalities on chip using electromagnetic (EM) waveguides is challenging due to their inherently high propagation velocity, which leads to large device footprints and limited achievable delay.

Acoustic waves, by contrast, travel roughly five orders of magnitude more slowly than EM waves, enabling lithographically defined wavelength-scale features and substantial delay at RF frequencies \cite{Jack_PnIC_2025}. As a result, piezoelectric acoustic devices have long been used for delay lines \cite{Ruochen_LNDL_2019,giribaldi_low_2023,Minghuang_LNOIDL_2020}, transversal filters \cite{mitsui_design_2025}, correlators \cite{Ghosh_correlators_2018}, convolvers, other analog signal-processing components \cite{Camp_SigProc_2012,guida_compact_2024} and quantum applications\cite{Krenner_2026}. Recent work has also demonstrated the feasibility of using acoustic delay lines directly in time-domain SIC architectures for IBFD systems, highlighting their potential for compact and broadband interference cancellation \cite{Jack_IBFD_2025}. However, although acoustic propagation losses are generally lower than those of EM structures, devices targeting large TBP—such as those required for multi-tap SIC networks—still experience significant cumulative attenuation. This loss fundamentally limits the achievable cancellation depth, dynamic range, and scalability of passive acoustic SIC implementations, motivating the incorporation of in-line amplification directly within the acoustic domain.

The acoustoelectric (AE) effect \cite{kino_AE_1976,Hutson_AETheory_1962} provides a compelling mechanism for such amplification. When the drift velocity of carriers in a biased semiconductor exceeds the acoustic phase velocity, energy is transferred from the carriers to the wave, resulting in net gain. The AE effect is inherently nonreciprocal: forward-propagating waves experience amplification, whereas reverse waves are attenuated. These characteristics make AE-enabled delay lines promising for loss-compensated SIC, large-TBP delay synthesis, and nonreciprocal RF front-end components required in IBFD systems. Efficient AE amplification requires pairing a high-electromechanical-coupling ($K^2$) piezoelectric substrate with a semiconductor exhibiting appropriate conductivity and high carrier mobility \cite{kino_theory_1971}. While piezoelectric semiconductors such as CdS \cite{White_CdS_1966}, GaAs \cite{Ludvik_GaAs_1972}, and GaN \cite{Ghosh_GaN_2019} are a compelling choice for AE integration, these materials offer limited $K^2$ and constrain achievable gain. Previous AE amplifiers often employed separated-medium architectures with a sub-micron air gap between the piezoelectric and semiconductor layers \cite{kino_theory_1971}. Although such devices demonstrate AE gain, the required alignment and packaging steps are challenging and limit scalability. Other heterogeneous material amplifiers have included direct deposition of the semiconductor such as InSb directly on LiNbO$_3$ \cite{10.1063/1.1653677}, and later hybrids that incorporate gating of two-dimensional electron gases such as GaAs on LiNbO$_3$ for strong SAW--carrier coupling with tunable conductivity \cite{10.1063/1.122400}. In contrast, modern layer-transfer techniques—such as direct wafer bonding—enable strong acoustic–electronic coupling through a thin dielectric interface \cite{Hackett_YXLN_2019,ghosh_experimental_2022}. These approaches simplify fabrication, allow integration with non-lattice-matched substrates, and are well suited for interfacing to high-coupling acoustic modes in thin-film lithium niobate.

Beyond these material systems, two state-of-the-art (SOA) material platforms have recently emerged for high-performance acoustoelectric amplifiers. The first approach combines InGaAs with high-coupling piezoelectric thin films such as scandium-doped aluminum nitride (ScAlN)\cite{Hackett_ScOSiC_2024}, leveraging both the high carrier mobility of III–V semiconductors and the enhanced piezoelectric coefficients of ScAlN to support strong acoustoelectric interaction at relatively high carrier concentrations. However, the sputtering-based deposition of ScAlN often leads to significant variability in surface roughness and crystalline quality\cite{Kapil_ScAlN_2026}, which in turn degrades interfacial uniformity and makes it challenging to achieve consistently high-yield wafer bonding across large areas. The second approach employs InGaAs integrated on lithium niobate on silicon (LN-on-Si) platforms \cite{hackett_non-reciprocal_2023}, which offers improved acoustic confinement and mechanical robustness compared to bulk piezoelectric substrates. Nevertheless, fundamental substrate-induced limitations restrict the achievable electro-acoustic coupling efficiency and hence the attainable acoustoelectric gain.

A more effective strategy is to combine a strong-coupling substrate—such as lithium niobate on insulator (LNOI) \cite{Hus_LNOI_coupling_2020, Hsu_LNOI_2021, Hsu_LNOI_gold_2024,hsu_thin-film_2021}—with a high-mobility semiconductor such as InGaAs \cite{Marin_InGaAsMob_2015}. In this work, we integrate metal–organic chemical vapor deposition (MOCVD)–grown InGaAs onto X-cut LNOI chips using an Al$_2$O$_3$-mediated wafer bonding process. A schematic illustration is shown in Fig.~\ref{fig:schem}(a). The platform leverages the high electron mobility of III–V semiconductors and the strong electromechanical coupling of LNOI to support efficient acoustoelectric interaction. Measurements on a single device are compared against an analytical AE gain model to establish the role of mode-dependent $K^2$ in determining the achievable gain, and reveal routes to broaden operating bandwidths and support higher-frequency operation. Hall-effect characterization of the transferred InGaAs is used as a quantitative diagnostic to connect the measured gain to the underlying material properties of the channel.

Beyond electromechanical coupling, we address a reliability failure specific to bare-channel III--V AE devices: the unprotected InGaAs surface oxidizes in ambient air and loses its AE response within weeks of fabrication. We demonstrate that an epitaxial InP barrier or an ALD Al$_2$O$_3$ layer enables stable operation, and analyze how each layer reshapes the piezoelectric field seen by the channel.

Finally, the underlying substrate also shapes the spatial distribution of the piezoelectrically induced field that mediates the acoustoelectric interaction: the buried, low-permittivity SiO$_2$ layer of the LNOI platform suppresses vertical field leakage and confines the piezoelectric potential near the lithium niobate surface, so that a larger fraction penetrates the overlying InGaAs channel \cite{Hwang_Coupling_2024}. Finite-element simulations of this field distribution are shown in Fig.~\ref{fig:schem}(b),(c), and the confinement advantage of the buried oxide is examined quantitatively against a LN-on-Si reference in Section~\ref{subsec:hall}. The following sections present the device theory, the growth, bonding and fabrication process, and the experimental results and future work.
\section{Theory}

The InGaAs-on-LNOI acoustic wave amplifier operates analogously to the separated-medium AE amplifier described by Kino and Reeder \cite{kino_theory_1971}. In the bonded structure used here, the thin Al$_2$O$_3$/InP dielectric stack replaces the original air gap, enhancing the coupling between the surface acoustic wave (SAW) electric fields and mobile carriers in the semiconductor. As illustrated in Fig.~\ref{fig:schem}, SAWs launched by the interdigital transducers (IDTs) propagate across the InGaAs mesa. The accompanying electric potential periodically modulates the carrier density, while an applied DC bias establishes a drift velocity~$v_d$. When $v_d$ exceeds the acoustic phase velocity $v_a$, the carrier–phonon interaction transfers energy to the wave, resulting in net amplification. This process is intrinsically nonreciprocal: waves traveling in the direction of carrier drift are amplified, whereas those propagating in the opposite direction are attenuated.

To model this interaction, we employ a gain formulation \cite{ghosh_acoustic_2022} that extends the classical AE theory to include realistic device geometries—specifically, thin semiconductor layers separated from the piezoelectric substrate by a finite dielectric bonding layer. The AE gain coefficient is expressed as

\begin{equation}
\alpha = \left(\frac{v_d}{v_a}-1\right)\cdot
\frac{G_0\,\omega\,\tanh(\beta d)/\beta d}
{\left(\frac{v_d}{v_a}-1\right)^2
+ C_0\left(1+\frac{\omega}{\omega_a}\right)\tanh(\beta d)/\beta d},
\label{eq:gain}
\end{equation}

where $v_d = \mu E$ is the carrier drift velocity under applied electric field, $v_a$ is the acoustic phase velocity, $\beta = \omega/v_a$ is the acoustic wavenumber, and $d$ is the semiconductor thickness. The material-dependent coefficients are

\begin{equation}
G_0 = \frac{\sigma d K^2}{2\varepsilon_p v_a}, \qquad
C_0 = \frac{\sigma d}{\varepsilon_p v_a}, \qquad
\omega_a = \frac{\varepsilon_g v_a}{\varepsilon_p h},
\label{eq:constants}
\end{equation}

where $\sigma$ is the semiconductor conductivity, $\varepsilon_p$ and $\varepsilon_g$ are the permittivities of the piezoelectric substrate and bonding dielectric, respectively, and $h$ is the thickness of the dielectric gap.

The frequency and coupling dependencies follow directly from Eqs.~\ref{eq:gain}–\ref{eq:constants}. The numerator increases linearly with operating frequency ($\propto \omega$), reflecting the faster interaction between carriers and the acoustic field at higher frequencies. The denominator introduces a frequency-dependent damping term of the form $(1+\omega/\omega_a)^2$, which arises from dielectric relaxation across the bonding layer. At low frequencies ($\omega \ll \omega_a$), carriers can follow the acoustic potential quasi-adiabatically, yielding an approximately linear rise in gain with $\omega$. At higher frequencies, however, the increasing phase lag between the carrier motion and the SAW potential suppresses energy transfer, establishing an optimal frequency band for maximum gain.

The electromechanical coupling coefficient $K^2$ enters through $G_0$, which scales proportionally with $K^2$. Since $K^2$ describes the efficiency of electromechanical energy conversion, larger values directly enhance the attainable AE gain. Consequently, SH-type modes supported in thin-film lithium niobate—with their inherently high $K^2$—offer substantially greater amplification potential than Rayleigh-type modes with weaker coupling.
\section{Growth, Bonding and Fabrication}

\begin{figure*}
\centering
\includegraphics[width = 1\textwidth]{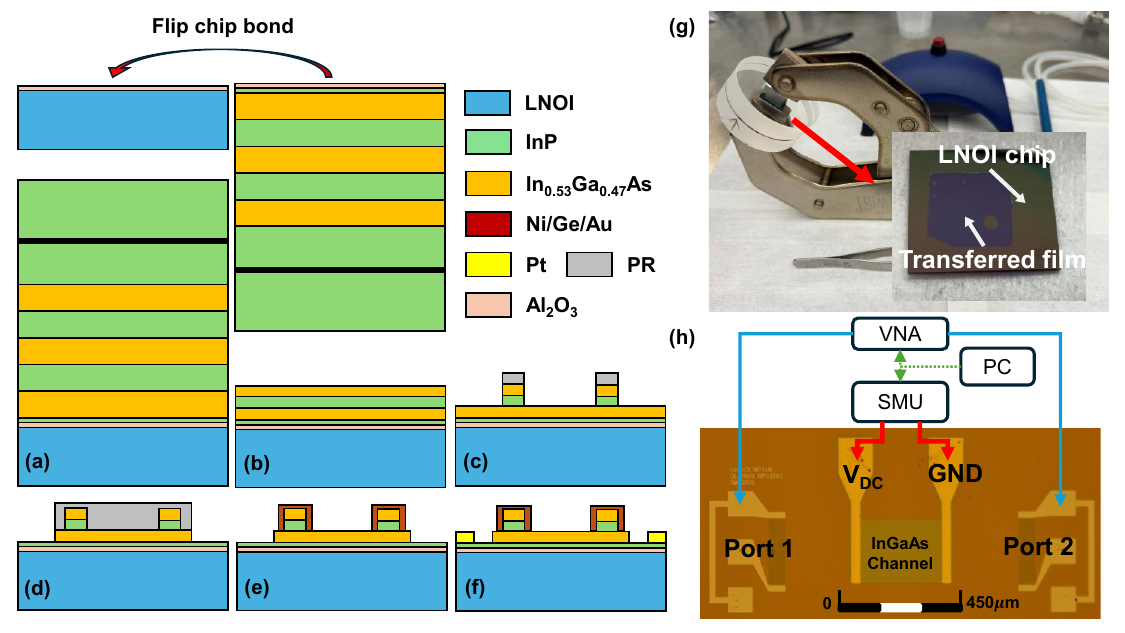}
\caption{\label{fig:wide} Fabrication process flow and measurement configuration of the InGaAs/LNOI acoustoelectric device. (a) Wafer bonding using an Al$_2$O$_3$ adhesion layer. (b) Handle substrate removal by selective wet etching. (c) Patterning of the InGaAs contact layer. (d) Channel mesa definition by wet etching. (e) Ohmic contact metal deposition followed by rapid thermal annealing. (f) IDT deposition and lift-off. (g) Photograph of the transferred thin film on the LNOI chip using a mechanical clamp. (h) Optical image of the fabricated device and experimental measurement setup for the acoustoelectric delay line, showing the electrical bias applied to the InGaAs channel and RF probing configuration.}
\end{figure*}

The InGaAs/InP heterostructure was grown on (100) InP substrates by metalorganic chemical vapor deposition (MOCVD) in an Aixtron EpiLab II reactor with a close-coupled showerhead geometry. A reactor pressure of 100 mBar, growth temperature of 615 °C (wafer surface), and V/III ratio of 80 for InP and 35 for InGaAs were used. The target layer structure is shown in Table~\ref{tab:epi_layers}.

\begin{table}[h]
\centering
\caption{Epitaxial layer structure of the InGaAs/InP heterostructure.}
\label{tab:epi_layers}
\begin{tabular}{llcc}
\hline
Layer Name     & Material                          & Thickness (nm) & Doping (cm$^{-3}$)  \\
\hline
Barrier        & InP                               & 5              & NID                  \\
Channel        & In$_{0.53}$Ga$_{0.47}$As          & 100            & NID                  \\
Barrier        & InP                               & 20             & NID                  \\
Ohmic Contact  & In$_{0.53}$Ga$_{0.47}$As          & 50             & $5 \times 10^{18}$   \\
Etch Stop      & InP                               & 100            & NID                  \\
Etch Stop      & In$_{0.53}$Ga$_{0.47}$As          & 100            & NID                  \\
Substrate      & InP                               & --             & --                   \\
\hline
\end{tabular}
NID = Non-Intentionally Doped
\end{table}

To integrate the high-mobility InGaAs channel with a piezoelectric substrate supporting strong electromechanical coupling, we developed a low-temperature, Al$_2$O$_3$-mediated wafer bonding process \cite{tian2026ifcs} that heterogeneously attaches the InGaAs layer onto X-cut LNOI. This approach circumvents the substantial lattice mismatch that prevents direct epitaxy on LiNbO$_3$ \cite{chae_monolithic_2024}. Surface preparation was performed prior to bonding to ensure atomically smooth and contaminant-free interfaces. The LNOI chip was cleaned in an RCA-1 solution (NH$_4$OH:H$_2$O$_2$:H$_2$O = 1:1:5) at 85°C for 45 min. Atomic force microscopy (AFM) measurements confirmed an RMS roughness of $\sim$0.4 nm after cleaning, which is within the acceptable range for successful direct bonding. The InGaAs wafer was diced into $\sim$1cm$^2$ dies and sequentially cleaned with acetone and isopropanol, followed by gentle Q-tip wiping to remove residual particles. A 4 nm Al$_2$O$_3$ layer was then deposited on both the InGaAs and LNOI surfaces via atomic layer deposition (ALD) using trimethylaluminum and H$_2$O precursors. The dies were aligned and brought into face-to-face contact using a mechanical C-clamp shown in Fig.~\ref{fig:wide}(g), and held at room temperature for 24 hours to establish pre-bonding. A subsequent anneal at 200°C for 24 hours in ambient air promoted covalent bond formation across the Al$_2$O$_3$ interface.

After bonding, the InP handle layer was removed using a selective wet etch (HCl:H$_2$O = 3:1), leaving behind the InGaAs/InP etch-stop structure. A second etch using H$_2$SO$_4$:H$_2$O$_2$:H$_2$O (1:1:10) selectively removed the InGaAs etch-stop layer, revealing the 50 nm highly doped InGaAs ohmic layer. Residual InP barrier layers were removed in dilute HCl to expose the active InGaAs channel. This bonding and film-transfer process provides a scalable route for integrating high-quality semiconductor layers with strong piezoelectric substrates, enabling efficient acoustoelectric amplification on a compact platform.

Figure~\ref{fig:wide}(b)–(f) shows the post-bonding microfabrication flow used to realize the AE delay lines on the LNOI platform. The process begins with defining the InGaAs mesa by patterning the top highly doped InGaAs layer and selectively etching it using an H$_2$SO$_4$:H$_2$O$_2$:H$_2$O (1:1:10) solution, followed by removal of the underlying InP barrier in an HCl:H$_2$O (3:1) etchant (Fig.~\ref{fig:wide}(c)). A second lithography step patterns the active channel region, which is then etched to form the lateral boundaries of the AE interaction region (Fig.~\ref{fig:wide}(d)). Ohmic contacts to the InGaAs are formed by depositing a Ni (10 nm)/Ge (25 nm)/Au (100 nm) metal stack through electron-beam evaporation, followed by lift-off. The contacts are annealed at 400°C for 1 min in a rapid thermal annealer to obtain low-resistance electrical access (Fig.~\ref{fig:wide}(d),(e)). Finally, platinum IDTs are patterned using a separate lift-off process (Fig.~\ref{fig:wide}(f)), which completes the definition of the AE delay lines.
\section{Experiments}
\begin{figure*}
\centering
\includegraphics[width = 0.93\textwidth]{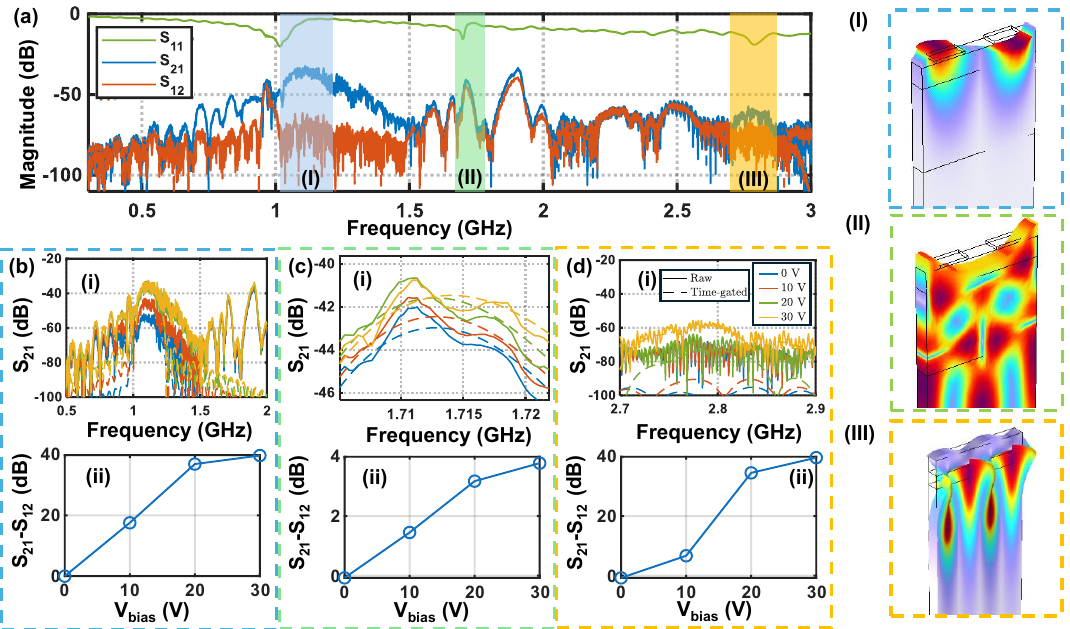}
\caption{\label{fig:mode} Mode-resolved acoustoelectric delay line characterization. (a) Raw transmission spectra measured under a $30~\mathrm{V}$ DC bias for a $3~\mu\mathrm{m}$ wavelength device with a $250~\mu\mathrm{m}$ acoustoelectric channel length, revealing three distinct frequency regions corresponding to different acoustic modes, as analyzed below.(b–d) Bias-dependent measurements for the three gated modes. For each mode, (i) raw and gated forward transmission $S_{21}$ under different DC bias voltages and (ii) the extracted maximum non-reciprocity $S_{21}-S_{12}$ as a function of bias voltage.}
\end{figure*}

\begin{table*}
\centering
\caption{\label{tab:table1}Simulated coupling coefficients and calculated acoustic Non-reciprocity (NR) corresponding to modes shown in Fig. \ref{fig:mode} and Fig. \ref{fig:CalcMode}}
\begin{tabular}{cccccc}
No. & Mode & \textit{K}$^2$ (\%) & Freq (GHz) & Calc. NR (dB/mm) & Mea. NR (dB/mm)\\
\hline
I   & SH-SAW           & 21.8 & 1.11 & 487 & 160\\
II  & LL-SAW           & 3.94 & 1.71 & 123 & 19.1\\
III & 2$^{nd}$ order SH-SAW & 13.9 & 2.79 & 648 & 174\\
\end{tabular}
\end{table*}
\begin{figure}
\includegraphics[width = 0.47\textwidth]{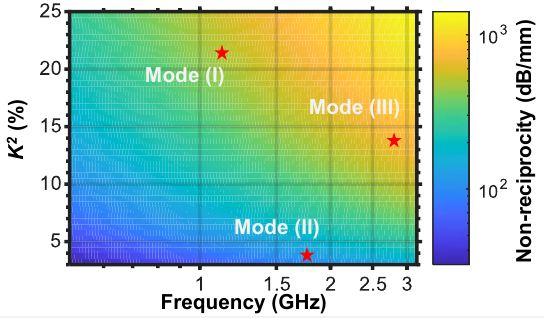}
\caption{\label{fig:CalcMode} Theoretical non-reciprocity mode map based on the analytical model. Calculated acoustoelectric non-reciprocity (color scale, dB/mm) as a function of frequency and effective electromechanical coupling coefficient $K^2$ obtained from the preceding theoretical formulation. 
}
\end{figure}

The AE gain structures were characterized by measuring their S-parameters using a vector network analyzer (VNA, Keysight P5008A) while simultaneously applying a DC bias to the InGaAs channel through the ohmic contacts. The DC bias was supplied using a source-measurement unit (SMU, Keysight B2902B). A schematic of the measurement configuration is shown in Fig.~\ref{fig:wide}(h). During each measurement, the DC bias was held constant as the S-parameter sweep was performed over the frequency range of interest, and the procedure was repeated for successive bias levels.

\subsection{AE delay line mode analysis}

This subsection establishes the feasibility of AE operation on the InGaAs-on-LNOI platform and identifies the acoustic modes it supports; the results reported here are an initial demonstration rather than a measure of the ultimate device performance. AE gain structures were fabricated with varying interaction lengths, and an example device together with the experimental setup is shown in Fig.~\ref{fig:wide}(h). Characterization was performed on an as-fabricated, unpassivated device with a 3~$\mu$m wavelength, examined within days of fabrication before ambient surface oxidation set in (Section~\ref{subsec:passivation}). Three distinct acoustic modes were examined: the fundamental shear-horizontal SAW (SH-SAW, mode I), a longitudinal-leaky SAW (LL-SAW, mode II), and the second-order SH-SAW (mode III). The zero-bias insertion loss originates partly from the transducer design---conventional bidirectional IDTs without SPUDT electrodes or impedance matching---rather than from the AE interaction; a bare-LNOI line of the same design shows the same passive response in simulation and measurement (Supplementary Fig. S1).

Fig.~\ref{fig:mode}(a) shows a raw transmission spectra measured under a $30~\mathrm{V}$ DC bias for a $3~\mu\mathrm{m}$ wavelength device with a $250~\mu\mathrm{m}$ acoustoelectric channel length, revealing three distinct frequency regions corresponding to different acoustic modes. Figures~\ref{fig:mode}(b--d) summarize the AE response for each mode. In panel (i), the measured S$_{21}$ spectra under different DC biases are shown, with the RF excitation power set to --5~dBm. Frequency- and time-gating were applied to isolate the desired mode and to suppress feedthrough and triple-transit reflections \cite{rf_measure_handbook}. Panel (ii) displays the peak difference between S$_{21}$ and S$_{12}$, quantifying the nonreciprocity introduced by AE gain.

The nonreciprocity increases monotonically with applied bias and approaches saturation near 30~V. Peak nonreciprocal contrasts (S$_{21}$--S$_{12}$ for time-gated data) of approximately 160~dB/mm, 19.1~dB/mm, and 174~dB/mm were measured for modes I, II, and III, respectively (Table~\ref{tab:table1}). The substantial variation in gain among the three modes reflects the dependence of AE amplification on both frequency and electromechanical coupling. Figure~\ref{fig:CalcMode} shows the theoretical gain map as a function of $K^{2}$ and frequency, with the three experimentally observed modes indicated. The extracted coupling coefficients and theoretical gain values are summarized in Table~\ref{tab:table1}, confirming that modes I and III possess significantly larger $K^{2}$ and therefore higher gain than mode II, consistent with experimental observations.

According to Eq.~(\ref{eq:gain}), AE gain increases with both carrier mobility and electromechanical coupling. Higher-order modes with non-negligible $K^{2}$ can therefore provide enhanced gain~\cite{Kramer_57GHz_2022}. While SH-SAW modes on X-cut LNOI are only weakly dispersive, their electromechanical coupling varies strongly with wavelength, leading to substantial mode-to-mode variations in achievable AE gain. Practical implementation thus requires careful wavelength and modal engineering to leverage the high-$K^{2}$ regimes while maintaining robust device performance. Table~\ref{tab:table1} confirms that the measured non-reciprocity follows the $K^{2}$--frequency trend predicted by the AE gain model. The following two subsections examine the surface stability and the transferred-film material properties of these devices.

\subsection{Surface Stability and Passivation}

\label{subsec:passivation}

\begin{figure}
\includegraphics[width = 0.483\textwidth]{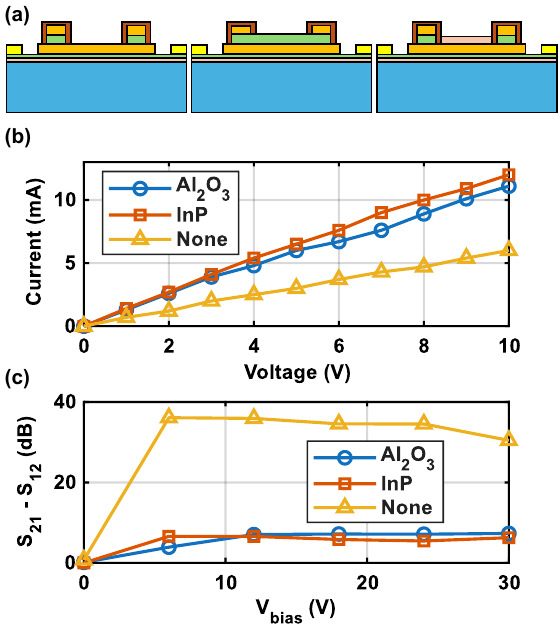}
\caption{\label{fig:Passivation} Impact of surface passivation on electrical transport and acoustoelectric performance. (a) Cross-sectional schematics of the device with different passivation conditions. (b) Current–voltage ($I$–$V$) characteristics of a $200~\mu\mathrm{m}$-long InGaAs channel under different passivation layers. (c) Extracted non-reciprocity ($S_{21}-S_{12}$) as a function of DC bias voltage for the corresponding passivation conditions.}
\end{figure}

An important practical consideration for InGaAs-based AE devices is the long-term stability of the active surface. When measured within days of fabrication, unpassivated devices exhibit strong AE non-reciprocity, as shown by the ``None'' curve in Fig.~\ref{fig:Passivation}(c). Over the following weeks of storage in ambient laboratory air however, the measured non-reciprocity decreases monotonically and ultimately falls below the noise floor of the measurement, despite no change in DC channel resistance large enough to account for the loss of AE response by transport considerations alone. Brief immersion of the aged device in dilute NH$_4$OH, followed by a deionized water rinse, restores the AE non-reciprocity to a level comparable to its original post-fabrication value. The reversibility of this degradation under a chemical treatment that selectively removes native oxides and arsenic-rich surface compounds from InGaAs identifies surface oxidation, rather than a bulk material change, as the underlying mechanism. The progressive growth of a non-uniform native oxide on the bare InGaAs introduces a high density of interface trap states and a screening dielectric layer that suppresses the piezoelectric potential reaching the mobile carriers, eliminating the conditions required for AE interaction.

This observation motivates the use of a deposited passivation layer to physically isolate the InGaAs surface from ambient oxygen and water vapor and thereby prevent the formation of the native oxide. Two passivation strategies are evaluated: a 20~nm InP barrier preserved from the epitaxial stack, and a 20~nm ALD-deposited Al$_2$O$_3$ layer applied directly to the exposed InGaAs surface after channel formation. The corresponding cross-sections are shown in Fig.~\ref{fig:Passivation}(a). Introducing any such passivation layer comes with a fundamental trade-off for AE devices: the additional dielectric increases the effective gap between the piezoelectric and the carrier channel, redistributing the piezoelectric potential away from the carrier-bearing region and lowering the peak attainable non-reciprocity for a fresh device. This trade-off is visible in Fig.~\ref{fig:Passivation}(c), where both passivated devices exhibit lower fresh-state non-reciprocity than the unpassivated control. The advantage of passivation is therefore not enhancement of the initial AE response but preservation of that response over device lifetime: passivated devices retain stable AE non-reciprocity over the timescales relevant for system-level deployment, whereas the unpassivated control loses its functionality within weeks. Among the two schemes investigated, both achieve adequate environmental protection; the choice between them is driven by integration considerations rather than by AE performance alone.

The I--V characteristics in Fig.~\ref{fig:Passivation}(b) further support this interpretation. The unpassivated channel shows reduced current at a given bias relative to either passivated structure, consistent with surface depletion induced by trap states and partial native-oxide formation between fabrication and measurement. Both passivated channels exhibit linear, near-ideal ohmic behavior over the measured bias range, indicating that the passivation also stabilizes the metal--semiconductor contact region against ambient conditions.

\subsection{Hall characterization and gain limit diagnostic}

\label{subsec:hall}

To connect the measured AE non-reciprocity to the underlying material properties, Hall-effect measurements were performed on the transferred InGaAs film. Measurements were conducted in a Lake Shore 8400 system in the van der Pauw configuration at room temperature, with a 0.3~T field, 1~$\mu$A current, and unity Hall factor, averaged over four geometries; a preceding ohmic-contact check gave linear $I$--$V$ characteristics. Reliable data could not be obtained on the unpassivated reference die due to the rapid surface evolution discussed in Section~\ref{subsec:passivation}. For the passivated samples, Hall measurements were performed on co-located test structures from the same chip as the AE devices. The InP-passivated structure yields $n \approx 1.1\times10^{17}~\mathrm{cm^{-3}}$ and $\mu \approx 8760~\mathrm{cm^{2}/V\cdot s}$, while the Al$_2$O$_3$-passivated structure yields $n \approx 9.0\times10^{16}~\mathrm{cm^{-3}}$ and $\mu \approx 7880~\mathrm{cm^{2}/V\cdot s}$. The mobilities are typical of high-quality In$_{0.53}$Ga$_{0.47}$As at this doping level, confirming that the transferred film has not suffered significant transport degradation. The carrier concentrations, however, are roughly two orders of magnitude above the design target of $\sim 3\times10^{15}~\mathrm{cm^{-3}}$. We attribute this elevated background doping to unintentional contamination in the growth chamber during epitaxy of the nominally undoped channel, rather than to a property of the transferred film or the bonding process. Since the InP barrier is epitaxial while the Al$_2$O$_3$ is deposited in a separate ALD step, the agreement of the two extracted concentrations to within an order of magnitude indicates that this doping level is set during growth as a bulk property of the channel, not by the surface layer. It is thus representative of the unpassivated device as well, whose bulk doping is the same and whose Hall data are unavailable only because surface oxidation impairs their measurability. Inserting the Hall-extracted $n$ and $\mu$ of each passivated device into Eq.~(\ref{eq:gain}) reproduces the non-reciprocity measured on that same device to within order of magnitude (32~dB/mm, fundamental mode); the design-density values in Table~\ref{tab:table1} (487, 123, 648~dB/mm) are the ideal upper bounds once the doping is corrected. We therefore treat this as a consistency check that identifies excess carrier density as the dominant limitation.

\begin{figure}
\includegraphics[width = 0.483\textwidth]{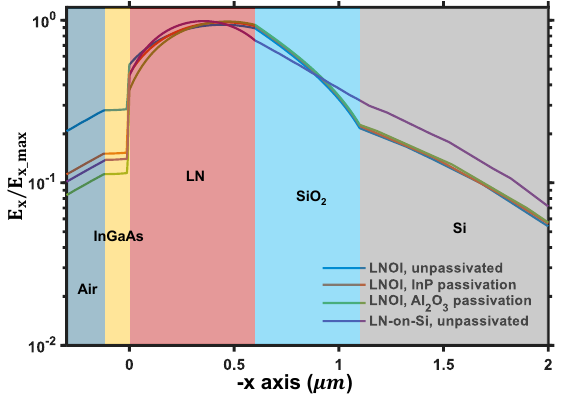}
\caption{\label{fig:Ecompare} Simulated normalized lateral electric field ($E_x/E_{x,\max}$) along the vertical cutline for LNOI with no passivation, InP passivation, Al$_2$O$_3$ passivation, and a LN-on-Si reference. All simulations assume an ideal intrinsic InGaAs channel (carrier concentration $3\times10^{15}~\mathrm{cm^{-3}}$, mobility $8740~\mathrm{cm^{2}/V\cdot s}$, conductivity $420~\mathrm{S/m}$) representing the design target rather than the actual material parameters. The colored background indicates the LNOI layer stack (air, InGaAs, LN, SiO$_2$, Si).}
\end{figure}

Passivation necessarily shifts part of the piezoelectric field off the channel, so a reduced channel field is expected for both schemes and is not specific to either. More importantly, the FEM simulation in Fig.~\ref{fig:Ecompare} highlights a structural advantage of the LNOI stack itself: the buried SiO$_2$ confines the piezoelectric field near the lithium niobate surface and keeps a large fraction of it in the active region, whereas the LN-on-Si reference shows markedly weaker confinement. The strong field confinement of the LNOI stack helps compensate for the field reduction that passivation introduces, so that a usable channel field---and hence acoustoelectric response---is retained even in the passivated, deployable configuration.

The platform-level $K^{2}$ and carrier mobility are consistent with expectations, leaving the channel doping as the only material parameter limiting present performance. The elevated conductivity also adds passive acoustic loss, as the mobile carriers dissipate acoustic energy. Bonding the unpassivated InGaAs lowers S$_{21}$ by $\sim$25~dB across the mode bands (Supplementary Fig. S2). Reducing the doping toward the design value through growth and chamber-conditioning optimization should therefore raise the achievable gain and lower the zero-bias insertion loss.
\section{Conclusion}

We have demonstrated SH-SAW acoustoelectric amplification on an epitaxial InGaAs / X-cut LNOI heterostructure formed by Al$_2$O$_3$-mediated wafer bonding, achieving a stable, deployable fundamental-mode non-reciprocity of 32~dB/mm in the passivated configuration at 30~V bias (64~mW DC power), while the unpassivated devices reach up to 174~dB/mm across modes spanning $\sim$1.1--2.8~GHz as an upper bound on the intrinsic AE capability. Mode-resolved characterization confirms the predicted dependence of AE gain on $K^2$ and frequency, with the second-order SH-SAW mode delivering the strongest normalized amplification. Hall-effect measurements identify a background carrier concentration approximately two orders of magnitude above the MOCVD design target, attributable to unintentional doping from chamber contamination during growth, as the principal limitation on absolute gain. This elevated carrier density is consistent, in magnitude and trend, with the measured non-reciprocity through the AE gain model, identifying excess doping as the dominant limitation rather than serving as a numerical validation of the framework. Equally important for moving this platform toward deployment, we show that the bare InGaAs channel is not stable in air---its AE non-reciprocity degrades within weeks---and that either an InP barrier or an ALD Al$_2$O$_3$ layer enables long-term operation. Because any such layer moves the piezoelectric field partly off the channel and lowers the fresh-state gain, passivation and gain must be optimized jointly rather than in isolation. With the epitaxial and passivation refinements outlined here, this platform offers a compact, low-power route toward multi-GHz non-reciprocal RF components and loss-compensated delay lines for in-band full-duplex transceivers and spectrum-efficient wireless front ends.
\section{Acknowledgments}
This work was supported by the DARPA Young Faculty Award (YFA) under Grant D24AP00005 and by the NSF CAREER Award 2340405.

\bibliographystyle{ieeetr}
\bibliography{bibliography}

\end{document}